\documentstyle[12pt]{article}
\begin{document}
\begin{center}
{\bf Symmetry Tests and Standard Model Backgrounds}\\
\medskip
Barry R. Holstein\\
Department of Physics and Astronomy\\
University of Massachusetts\\
Amherst, MA 01003\\
\medskip
{\bf Abstract}
\end{center}
\begin{quote}
Symmetry tests provide an important probe for the structure of elementary
particle interactions and for the validity of the standard model.  However,
it is pointed out that in the interpretation of such experiments one must
keep in mind that in many cases apparent "violations" of such tests are
actually the result of ordinary effects within the standard model.
\end{quote}

\begin{tabbing}
the quick brown fox jumped over the lazy dog\= the lazy dog \=          \kill
                               \> {\em God, thou great symmetry,}\\
                               \> {\em Who put a biting lust in me}\\
                               \> {\em From whence my sorrows spring,}\\
                               \> {\em For all the frittered days}\\
                               \> {\em That I have spent in shapeless ways}\\
                               \> {\em Give me one perfect thing.}\\
                               \>       \>{\em Anna Wickham}
\end{tabbing}

\section{Introduction}
The name Ernest Henley to me is inexorably linked with the idea of symmetry
tests.  Indeed I learned my first information about parity violation in
nuclei from his classic work on time reversal and parity\cite{1}
and it was reading this paper which
eventually led to my interest in and work in this field.  Of course, it is
not just Prof. Henley who has long been interested in the subject of
invariance.  Indeed, mankind has from the earliest days
been fascinated with the concept of symmetry.  The
Pythagorans considered the circle and sphere to be the most perfect of two
and three dimensional objects respectively because of their obvious radial
symmetry.  The planets were assumed to move in precise circular orbits and the
stars
were assumed to be situated in the heavenly spheres. It is during recent
decades that
symmetry studies in physics have had a rebirth, however.  The reason for this
has to do with the development of modern physics.  As long as physics was
focussed on its classical roots of mechanics and electrodynamics, for which
the exact laws of motion were known, the use of symmetry methods was,
strictly speaking, not
necessary.  Indeed most books on classical mechanics written during the
early part of the present century do not even mention the conservation of
angular momentum---they just solve the Kepler problem exactly.\cite{2}
However, with the advent of particle and nuclear
physics, wherein the underlying interactions and equations of motion are
complex
and unknown, the use of symmetries in order to flesh out the structure of
these fundamental interactions has become commonplace, especially after the
discovery of parity violation within the weak interaction in 1957.\cite{3}

In the interpretation of the results of such tests, however,
it must always be kept
in mind that apparent violations can be the result of quite pedestrian
effects within
the standard model rather than bona fide symmetry breaking.  A familiar
example is the mocking of T-violation by final state strong interactions.
Similar effects are expected for other symmetries and below we review the
size of these standard model "backgrounds" which can be expected in such
tests.  Knowledge of the size of such symmetry violation simulations is
essential in planning experiments since this represents a fundamental
and inescapable upper bound on the sensitivity of these measurements.

\section{Symmetry Tests in Electroweak Physics}

When as particle/nuclear physicists we write down the so-called standard
model of electroweak interactions, we simultaneously build in without
thinking a significant number of symmetry assumptions, each of which
is intrinsic to the standard electroweak model and is
subject to experimental test.

Because of space limitations I shall concentrate here on the low-energy and
light quark sector of the weak interaction, which then takes the form
\begin{eqnarray}
& &{\cal L}_{wk}=-{G\over \sqrt{2}}J^\mu J^\dagger_\mu \qquad {\rm
with}\nonumber\\
J_\mu &=&\sum_{e,\mu}\bar{\ell}\gamma_\mu(1+\gamma_5)\nu_\ell +
(\bar{d}\quad\bar{s})\gamma_\mu(1+\gamma_5)U_{KM}\left( \begin{array}{c}
u\\c\end{array}\right)
\end{eqnarray}
where $U_{KM}$ is the KM matrix in this sector.  The set of processes
allowed by this Lagrangian is a large one and includes:
a) leptonic: $\mu^-\rightarrow e^-\bar{\nu}_e\nu_\mu$,
b) semileptonic: $n,\Lambda\rightarrow
pe^-\bar{\nu}_e,\mu^-p\rightarrow n\nu_\mu$, and
c) nonleptonic: $np\rightarrow np,\Lambda\rightarrow p\pi^-$ reactions.
We begin by considering each symmetry in turn and the means by which it
can be tested.
Many such tests involve correlation measurements in nuclear beta decay.
It is then useful to summarize the set of such correlations, and we do so in
the
Appendix.

i) V,A Character: The form of the various beta-decay correlations in the
case of V,A coupling is given in many sources.\cite{4}  The deviations expected
in the presence of S,P or T interactions can be found in the classic work
of Jackson, Treiman and Wyld [JTW].\cite{5}  A classic test for the presence of
such
additional couplings is to look at the longitudinal polarization of the
outgoing electron/positron in a beta decay process.
An alternative tack in the case of Fermi transitions is to look for a
systematic dependence\cite{9d}
\begin{equation}
ft\propto (1-2b_F\gamma<{m_e\over E}>)\quad{\rm with}\quad \gamma=\sqrt
{1-\alpha^2Z^2}
\end{equation}
in the ft values of such decays.
Recently Adelberger has pointed out that the broadening of the proton peak in
delayed proton emission accompanying beta decay can also be used in order to
provide improved limits on S,T couplings.\cite{9c}  Averaging measurements such
as
these one finds rather strong limits\cite{7}---$b_F<0.005$---on
the absence of such non-V,A couplings

Finally, recent claims have been made that analysis of radiative pion
decay experiments---$\pi^-\rightarrow e^-\bar{\nu}_e\gamma$---performed at
Serpukov has provided evidence for the existence of a nonzero tensor
coupling.  However, the "measured"
number\cite{9a}---$F_T=-0.0066\pm 0.0023$---is in contradiction with the upper
bound---$F_T=0.0018\pm0.0017$---given by beta decay
measurements\cite{9b} and should I believe be discounted.

ii) Left-handedness:  In seeking evidence for the possible presence
of right-handed weak currents, again the best sensitivity comes from
examination
of possible deviations of experimental correlation measurements from the
values expected within a pure left-handed picture.  It is traditional to
parameterize the results in terms of\cite{9e}
\begin{equation}
\sigma={m_L^2\over m_R^2}\quad {\rm and}\quad \zeta=\mbox{L-R mixing angle}
\end{equation}

\vspace{2.5in}

Figure 1: Experimental limits on right-handed currents.

\medskip
\noindent Again, one of the best present limits arises from electron/positron
longitudinal
polarization measurements,\cite{6}
while another approach involves the use of the beta-polarization coefficient A
in superallowed decay as a probe of right handed effects.  In the latter case
the most stringent limit
is obtained in cases where the measured asymmetry is small such as
${}^{19}$Ne\cite{9f} or neutron beta decay.\cite{9g}
A third and powerful probe is provided by the Michel parameters in muon
decay.  Finally a new approach has been proposed by Quin and Gerard who pointed
out
the sensitivity of the nuclear-electron spin correlation function to the
presence of right-handed currents.\cite{9h}  A recent such experiment has been
performed
using ${}^{107}$In by Severijns et al.\cite{9i}
The present limits on the right-handed parameters $\eta ,\zeta$ are
summarized in Figure 1.

iii) CVC:  There exist two classic tests of CVC within nuclear beta decay.
The most familiar (and most precise) is the prediction of identical ${\cal F}t$
values for superallowed $0^+-0^+$ decays.  Here
\begin{equation}
{\cal F}t={2\pi^3\ln 2\over G_\mu^2m_e^5|V_{ud}|^2 a^2(0)}(1-{2\alpha\over \pi}
\ln {m_Z\over m_N}+\cdots )
\end{equation}
where $a(0)=\sqrt{2}$ is the CVC requirement on the Fermi matrix element.
Such an analysis has been
performed by many groups and all results agree at the $0.5\%$ level.\cite{7}
However, as
discussed shortly, it is important to know the result even better.
The second test involves the comparison of shape factors in mirror
$\beta^+,\beta^-$ decays, which are predicted in lowest order for the familiar
A=12 system\cite{9j} in terms of the measured M1 decay width from the
15.11 MeV ${}^{12}$C excited state to be
\begin{equation}
{dS_\pm\over dE}\approx \mp{4\over 3}{b\over Ac}{1\over m_N}\sim \mp 0.5\%/MeV
\end{equation}
The experimental number obtained by Lee, Mo
and Wu using an iron free magnetic spectrometer yields values in good
agreement with this prediction.\cite{9k}

iv) PCAC:  In the context of beta decay/muon capture PCAC makes two
predictions.  One is the Goldberger-Treiman relation\cite{9m}
\begin{equation}
F_\pi g_{\pi NN}=m_Ng_A
\end{equation}
relating strong and weak nucleon couplings.  The second is that the
induced pseudoscalar should have the size\cite{9n}
\begin{equation}
r_P={m_\mu\over 2m_N}g_P(q^2=-0.9m_\mu^2)=7.0
\end{equation}
in a muon capture process.

Unfortunately at the present time the precise validity of both predictions is
open to question In the case of the Goldberger-Treiman relation the problem
has primarily to do with the value of the strong $\pi -N$ coupling constant,
as will be discussed in the next section.  In the case of the induced
pseudoscalar the difficulty is that sensitivity is available only in muon
capture experiments and even then only at the cost of significant model
dependent assumptions.  Present results are
\begin{equation}
r_P=\left\{\begin{array}{ccc}
7.4\pm 2.0 & {}^3{\rm He} &\cite{9o}\\
6.5\pm 2.4 & {\rm p} &\cite{9p}\\
9.1\pm 1.7 & {}^{12}{\rm C} &\cite{9q}
\end{array}\right.
\end{equation}
Each of these results agrees then with the PCAC prediction, but the accuracy
is only at the 30\% level and could be substantially improved.

v) G-Invariance:  Using the quark model one finds that the usual polar-,
axial-vector currents satisfy the relations
\begin{equation}
GV_\mu G^{-1}=V_\mu ,\quad
GA_\mu G^{-1}=-A_\mu
\end{equation}
Second class currents were defined by Weinberg as being those having
opposite signs under the G-parity operation and are not present in the
standard model.\cite{9r}  Their absence has definite implications for weak
matrix elements.  The most general axial vector matrix element
between spin 1/2 systems has the form
\begin{equation}
<p|A_\mu|n>=\bar{u}(p_2)\left(g_1\gamma_\mu-i{g_2\over 2m_N}\sigma_{\mu\nu}
q^\nu+{g_3\over 2m_N}q_\mu\right)\gamma_5u(p_1)
\end{equation}
Here for an analog transition such as neutron decay, the absence of second
class currents requires $g_2=0$.\cite{9s}
 In the case of {\it nuclear} beta decay, the analog
of $g_2$ is the tensor form factor d and the absence of second class
currents requires that\cite{9s}
\begin{itemize}
\item [i)] $d=0$ for an analog transition, e.g. ${}^{19}Ne\rightarrow{}^{19}F$
\item[ii)] $d_{\beta^+}=d_{\beta^-}$ for mirror decay, e.g.
${}^{12}B\rightarrow
{}^{12}C\leftarrow{}^{12}N$
\end{itemize}
The tensor term may be measured via correlation experiments, and the best of
these measurements involves the alignment correlation in the A=12 mirror
system,\cite{9t} for which the present experimental number is\cite{9u}
\begin{equation}
d^{II}/b=-0.05\pm 0.13
\end{equation}
{\it i.e.} second class currents are ruled out at the level of 20\% of weak
magnetism---not a particularly precise limit.

vi) T-Invariance:  Neglecting final state interaction effects, time reversal
invariance requires that all amplitudes contributing to a process be relatively
real.  Thus one looks for T-violation by seeking a phase difference between
two or more multipoles with participate in a decay.  The most precise
experimental work has been done by measuring the D coefficient in beta
decay
\begin{equation}
D^{\rm exp}=\left\{\begin{array}{ccc}
(0.7\pm 6)\times 10^{-4}& {}^{19}{\rm Ne}&\cite{9v}\\
(-1.1\pm 1.7)\times 10^{-3} & {\rm n}&\cite{9w}
\end{array}\right.
\end{equation}
On the theoretical side a non-zero value of D can arise from a phase difference
between leading Fermi and Gamow-Teller terms or between leading and recoil
terms.\cite{9x}
An alternative approach is to measure the R coefficient, which is sensitive
to a possible phase difference between V,A and a tensor interaction\cite{5}
Present experimental numbers are consistent with zero but with sizable errors
\begin{equation}
{\rm Im}(G_T+G_T')=\left\{\begin{array}{ccc}
0.136\pm 0.091 & {}^{19}{\rm Ne}&\cite{9y} \\
0.024\pm 0.027 & {}^8{\rm Li}&\cite{9z}
\end{array}\right.
\end{equation}
However, since such experiments involve measurement of the electron
polarization,
they will never approach the statistical precision of the corresponding
measurements of the D correlation.

\section{Standard Model Backgrounds}

Two standard model symmetries have no corrections even when all
components of the standard model are considered.  These are CPT, whose validity
follows simply from the tenets demanded of any reasonable quantum field
theory, and the other is the left-hand nature of the weak interaction,
which is only modified by going outside the standard model.  All other
symmetries, however, are no longer exact when higher order effects
are included.

One of these is not particularly important, however.  In the case of the V-A
nature of the weak interaction, Higgs boson exchange can introduce
effective S,P components into the weak interaction.  However, in view of the
present mass limits on the Higgs, such effects should be negligible.

i) T-invariance:  In the case of T invariance, any real
effects arising from higher loop
graphs within the standard model are negligible since heavy quarks must
be involved.  However, as is well-known, T-violation can be simulated by
strong and electromagnetic final state interactions, which according to the
Fermi-Watson theorem, give rise to different phases for different multipole
amplitudes.  In the case of the very
precise D coefficient measurements, a quick glance at JTW indicates that
within the V-A picture there are no such corrections at leading order.
However, at the level of recoil this is no longer the case and one
finds from one-photon exchange\cite{9aa}
\begin{eqnarray}
D^{EM}&=&{1\over |a|^2+|c|^2}(\pm {Z\alpha E^2\over 4Mp}[\delta_{JJ'}
\left({J\over J+1}\right)^{1\over 2}
{\rm Re}a^*\left((b\mp c)(1+3{m^2\over E^2})\right.\nonumber\\
&-&\left.d(1-{m^2\over E^2})
\right)
-{1\over 2}{\gamma_{JJ'}\over J+1}{\rm Re}c^*(c\pm d\mp b)(3+{m^2\over E^2})]
+\cdots
\end{eqnarray}
For these superallowed transitions then the leading effect comes from
interference
between the Gamow-Teller and weak magnetism form factors leading to small but
non-negligible values
\begin{equation}
D^{EM}=\left\{ \begin{array}{cc}
2\times 10^{-4}& {}^{19}{\rm Ne} \\
2\times 10^{-5}& {\rm n}
\end{array}\right.
\end{equation}
In fact such experiments if done to this precision can be turned
around---measurement of the final state
interaction effect can be used as a probe of weak magnetism even
if no bona fide T-violating signal is detected

For other correlations the electromagnetic effect can arise at leading
(non-recoil)
order.  Thus one finds for the R coefficient
$R^{EM}/A \approx \alpha Zm/ p$\cite{5}
which can simulate T violation even at the $10^{-2}$ level in some cases.

ii)PCAC:  A careful look at the derivation of the Goldberger-Treiman relation
shows that it should read\cite{9n}
\begin{equation}
F_\pi g_{\pi NN}(q^2)=M_Ng_A(q^2)
\end{equation}
so that both the strong and weak couplings should be evaluated at the {\it
same}
value of momentum transfer.  Generally what is quoted and used, however,
are $g_A(0)$ and $g_{\pi NN}(m_\pi^2)$.  Thus a discrepancy is {\it expected}
for the Goldberger-Treiman relation, and this can be characterized in terms
of the quantity
\begin{equation}
\Delta_\pi =1-{M_Ng_A(0)\over F_\pi g_{\pi NN}(m_\pi^2)}
\end{equation}
for which one expects $\Delta_\pi\sim m_\pi^2/ 2m_\sigma^2\sim 0.015$
\cite{9cc}
The experimental size of the discrepancy is at present unclear, due to
uncertainty over the size of the pi-nucleon coupling constant.  The situation
is summarized in Table 1.

\begin{table}
\begin{center}
\begin{tabular}{|c|c|c|}\hline
  & $g^2/4\pi$ & $\Delta_\pi$\\ \hline
$\pi^\pm$ &13.31(27)\cite{aa}&0.002\\
 & 14.28(18)\cite{bb} & 0.043 \\ \hline
$\pi^0$&13.55(13)\cite{cc}&0.017\\
 &14.52(40)\cite{dd}&0.051\\ \hline
\end{tabular}
\end{center}
\caption{Experimental values of the pion-nucleon coupling constant and the
associated Goldberger-Treiman discrepancy.}
\end{table}

iii) G-invariance:  If we write the most general axial matrix element between
neutron and proton as in Eq. 10
then, since neutron and proton are members of a common isomultiplet,
G-invariance requires $g_2=0$.  Many precise measurements have attempted
to check this prediction.  However, as discussed above, since this structure
function is associated with recoil the experimental limits obtained thereby
are relatively weak---$g_2^{\rm exp}< 0.4$.

Within the standard model, one expects that $g_2$ should be nonvanishing
due to both electromagnetic effects and quark mass differences.  In particular
the latter can be estimated within a relativistic quark model, wherein
one finds\cite{9dd}
\begin{equation}
{g_2\over g_A}={\int d^3x r(u_u\ell_d-u_d\ell_u)\over
\int d^3x (u_uu_d-{1\over 3}\ell_u\ell_d)}
-{1\over 4}({m_n\over m_p}-{m_p\over m_n})
\end{equation}
For $\Delta S=0$ processes such as nuclear beta decay one finds $g_2/g_A\sim
10^{-3}$ so that the effect is essentially unmeasurable.  However, for $\Delta
S=1$ hyperon decays the strange quark mass is involved and one finds
$g_2/g_A\sim 0.3-0.4$, which should certainly be detectable.  Unfortunately
previous analyses have not been precise enough to see this effect.  In
fact usually $g_2$ is simply set to zero.  However, future work involving
correlation studies together with rate
measurements should be able to resolve this question

iv) SU(2), SU(3):  For the case of SU(2) violation a particularly illuminating
example involves $K_{\ell 3}$ decays
\begin{equation}
K^+\rightarrow\pi^0e^+\nu_e\quad{\rm and}\quad K_L\rightarrow\pi^-e^+\nu_e
\end{equation}
For the P-wave coupling in such decays, one would have in exact SU(2)
\begin{equation}
f_+^{K^+\pi^0}(0)/f_+^{K_L^0\pi^-}(0)=1
\end{equation}
and one might suspect little change in this result because of the Ademollo
Gatto theorem, which seems to assert that any violation of Eq. 20 must be
second order in symmetry breaking.\cite{9ee}  However, this is not the case. In
fact
because of $\eta -\pi^0$ mixing one has
\begin{eqnarray}
\quad 1&-&|f_+^{K_L^0\pi^-}(0)|^2={\cal O}(\epsilon^2)\nonumber\\
{\rm but}\quad 1&-&{1\over 4}|f_+^{K^+\pi^0}(0)|^2-{3\over
4}|f_+^{K^+\eta^0}|^2=
{\cal O}(\epsilon)
\end{eqnarray}
Thus one predicts\cite{9ff}
\begin{equation}
f_+^{K^+\pi^0}(0)/f_+^{K_L^0\pi^-}(0)\approx 1+{3\over 4}{m_d-m_u\over m_s-
{1\over 2}(m_d+m_u)}=1.02
\end{equation}
which is in good agreement with the experimental number
\begin{equation}
f_+^{K^+\pi^0}(0)/f_+^{K_L^0\pi^-}(0)=1.029\pm 0.010
\end{equation}

In the case of SU(3) violation, it is interesting to examine semileptonic
hyperon decay---$\Lambda\rightarrow pe^-\bar{\nu}_e,\Sigma^-\rightarrow
ne^-\bar{\nu}_e,etc.$  Generally such decays are fit via the assumption of
SU(3) symmetry
\begin{equation}
<B_b|J^c_\mu|B_a>=\bar{u}_b(F_Vf_{abc}\gamma_\mu +(F_Af_{abc}
+D_Ad_{abc})\gamma_\mu\gamma_5)u_a
\end {equation}
and such fits are very good but certainly not perfect.  In fact there is
good evidence for SU(3) breaking from such fits if one compares the
experimental and theoretical predictions in the case of the $\Sigma-\Lambda$
transition.  Defining the SU(3) breaking parameter $\rho$ via
\begin{equation}
g_A^{\Sigma^-\Lambda}=\rho\sqrt{2\over 3}{D\over D+F}g_A^{np}
\end{equation}
we find $\rho=0.914\pm 0.022$ when the experimental value for $g_A^{\Sigma^-
\Lambda}$ and the fit value for $D,F$ are employed.\cite{9gg}  However, it
should be
kept in mind that the assumption $g_2=0$ was made in performing such fits.  In
any case it is necessary to understand such effects in {\it both} meson and
baryon
sectors in order to extract believable values of $V_{us}$.

v)CVC:  The importance of electromagnetic effects which violate the naive
prediction $a(0)=\sqrt{2}$ is critical and possibly lies at the origin of
possible KM matrix unitarity violating effects which have been reported by
some groups.  Thus including radiative and other effects one finds\cite{9hh}
\begin{equation}
{\cal F}t=ft(1+\Delta_{\beta}+\delta_r+{\alpha\over\pi}C_{NS})(1-\delta_c
+\Delta_c^Z)
\end{equation}
where here $\Delta_\beta , \delta_r$ are the usual radiative correction
factors, $C_{NS}$ is an axial current correction, $\delta_c$ is a valence
nucleon mismatch factor and $\Delta_c^Z$ is a term recently proposed by
Wilkinson to account for core nucleon mismatch.\cite{9ii}  Using the form
$\Delta_c^Z
\sim\gamma Z^{1.8}$ one can achieve a reasonably good fit to the $0^+-0^+$
decay ft values and use of this correction brings
about no violation of the KM unitarity condition
\begin{equation}
1-\sum_j|V_{uj}|^2=\left\{\begin{array}{ll}
0.0044(12) & {\rm without}\quad \Delta_c^Z\\
0.0008(12) &{\rm with}\quad \Delta_c^Z\end{array}\right.
\end{equation}
However, this simple phenomenological procedure is not substitute for a careful
theoretical analysis.

\section{Conclusion}

We have seen that use of the standard model in order to describe the
electroweak
interactions implicitly assumes the validity of a host of symmetries---CVC,
PCAC,
G-invariance, etc.  Each of these symmetries (except CPT) is subject to
experimental verification in low energy leptonic and semileptonic decays.
However, in many cases the standard model also provides the ultimate
"background" to such tests, in producing non-zero results whose origin
is not at all related to the symmetry violation which one is trying to probe.

\begin{center}
{\bf Appendix}
\end{center}
The standard notation for correlation parameters in nuclear beta decay was
given by Jackson, Treiman and Wyld and has the form\cite{5}
\begin{eqnarray}
d\Gamma&=&\Gamma_0(1+{{\bf p}\cdot{\bf p}_\nu\over EE_\nu}+{m\over E}b
+<{\bf J}>\cdot\left[{{\bf p}\over E}A+{{\bf p}_\nu\over E_\nu}B
+{{\bf p}\times {\bf p}_\nu\over EE_\nu}D\right]\nonumber\\
&+&<{\bf \sigma}>\cdot\left[{{\bf p}\over E}G+{<{\bf J}>\times{\bf p}\over E}
R\right]
+<J_iJ_j>\left[ \left({p_ip_j\over E^2}-{p^2\over 3E^2}\delta_{ij}
\right)H\right.\nonumber\\
&+&\left.\left({p_ip_{\nu j}\over EE_\nu}-{{\bf p}\cdot{\bf p}_\nu\over
3EE_\nu}
\delta_{ij}\right)K+\cdots\right])
\end{eqnarray}
Here the correlation parameters A,B,C, etc. are expressed in terms of the
nuclear form factors, which for a general allowed transition can be written
in the form\cite{9s}
\begin{eqnarray}
<\beta_{p_2}|V_\mu(0)&+&A_\mu(0)|\alpha_{p_1}>={1\over
2M}aP\cdot\ell\delta_{JJ'}
\delta_{MM'}-{i\over 4M}\epsilon_{ijk}(J'M'1k|JM)\nonumber\\
&\times&[2b\ell_iqj+i\epsilon_{ij\lambda\eta}\ell^\lambda(cP^\eta-dq^\eta)]+
\cdots
\end{eqnarray}
where J,J' are the spins of the parent and daughter nuclei respectively, and
M,M' represent the initial and final components of nuclear spin along some axis
of quantization. The four-vector quantity
$\ell^\mu=\bar{u}(p)\gamma^\mu(1+\gamma_5)v(k)$ is the lepton matrix
element and a,b,c,d represent reduced matrix elements.  Using standard notation
\begin{equation}
a=g_VM_F,\qquad c=g_AM_{GT}
\end{equation}
where $M_F,M_{GT}$ are the Fermi, Gamow-Teller matrix elements respectively,
while b is the weak-magnetism contribution which, between nuclear analog
states would be given by
\begin{equation}
b=A({J+1\over J})^{1\over 2}M_F\mu_V
\end{equation}
where A is the mass number and $\mu_V$ is the isovector magnetic moment
measured
in terms of nuclear magnetons.  The coefficient d, the induced tensor, is
uniquely correlated with the existence of second class currents if
$\alpha,\beta$
are isotopic analogs. On the other hand, if $\alpha,\beta$ are not members of
a common isotopic multiplet, the existence of d is not forbidden by G-parity
considerations and even receives a contribution from first class currents in
the
nuclear impulse approximation.

\medskip

\begin{center}
{\bf Acknowledgement}
\end{center}

This work is supported in part by the National Science Foundation.

\end{document}